\documentstyle[epsf,12pt]{article}

\begin{document}

\title
{Universal invariant renormalization of supersymmetric Yang-Mills
theory.}

\author{A.A.Slavnov\thanks{E-mail:$slavnov@mi.ras.ru$}\\
{\small{\em Steklov Mathematical Institute,
117966, Gubkina, 8, Moscow, Russia and}}\\
{\small{\em Moscow State University, physical faculty,
department of theoretical physics,}}\\
{\small{\em $117234$, Moscow, Russia}}
\\
\\
and K.V.Stepanyantz\thanks{E-mail:$stepan@theor.phys.msu.su$}\\
{\small{\em Moscow State University, physical faculty,
department of theoretical physics,}}\\
{\small{\em $117234$, Moscow, Russia}}}

\maketitle

\begin{abstract}
A manifestly invariant renormalization scheme of $N=1$ nonabelian
supersymmetric gauge theories is proposed.
\end{abstract}

\sloppy


\section{Introduction}
\hspace{\parindent}

Construction of a manifestly supersymmetric and gauge invariant
renormalization procedure is a nontrivial and in some cases not
yet solved problem. Certainly, it is preferrable to have a calculation
scheme which preserves supersymmetry at any intermediate stage. However,
most popular gauge invariant regularizations like dimensional
regularization \cite{tHooftVeltman}, dimensional reduction \cite{Siegel}
or lattice regularization break supersymmetry \cite{JackJones}. Higher
covariant derivative regularization \cite{Slavnov,Bakeyev} can be easily
used in Abelian theories \cite{St1,3loop,Nishihara}. In principle, higher
derivative regularization is also applicable to non-Abelian theories
\cite{West_Paper}, but the calculations of quantum corrections are quite
involved. In this case it can be convenient to use non gauge invariant
regularization. For example, one can regularize the theory by adding
higher derivative term with ordinary derivatives, instead of the
covariant ones for the Yang-Mills field which simplifies the
calculations considerably. Such regularization preserves supersymmetry,
but breaks gauge invariance, which should be restored by a proper choice
of a renormalization scheme. For example, in the framework of algebraic
renormalization it is done by tuning finite counterterms to provide
relevant Slavnov-Taylor Identities (STI) for the renormalized Green
functions. This method was applied successfully to SUSY gauge theories
in the papers \cite{PS1,PS2}, where the invariant renormalizability of
$N=1$ non-Abelian SUSY gauge models was proven in the framework of
algebraic renormalization. However a practical implementation of the
algebraic renormalization is rather cumbersome as the procedure requires
a tuning of a large number of (noninvariant) counterterms.

Recently a new method of invariant renormalization was proposed
\cite{Sl3,Sl4}, which provides automatically the renormalized Green
functions satisfying STI. This method was generalized to sypersymmetric
electrodynamics in \cite{SlSt}. However, in order to apply it to
non-Abelian supersymmetric theories it is necessary to solve some problems,
caused by a more complicated structure of STI. This is done in the
present paper.

The paper is organized as follows:

In Section \ref{Section_SUSY_YM} we introduce the notations and
remind some information about supersymmetric Yang-Mills theory.
The universal invariant renomalization scheme for the model is
constructed in Section \ref{Section_Universal_Renormalization}.
This scheme is illustrated in Section \ref{Section_Example} by
calculation of the one-loop $\beta$-function with the simplified
(noninvariant) version of higher derivative regularization. The results
are discussed in the Conclusion.


\section{$N=1$ supersymmetric Yang-Mills theory.}
\hspace{\parindent}
\label{Section_SUSY_YM}

$N=1$ supersymmetric Yang-Mills theory may be described by the following
action: \footnote{In our notations the metric tensor in the Minkowski
space-time has the diagonal elements (1, -1, -1, -1).}

\begin{eqnarray}
&& S = \frac{1}{2 e^2}\mbox{Re}\,\mbox{tr}\int d^4x\ d^2\theta\
W_a C^{ab} W_b
+\nonumber\\
&& \qquad\quad
+ \frac{1}{16 e^2} \mbox{tr} \int d^4x\,d^4\theta \Bigg(
\Big(\bar c^+ -\bar c\Big)\,\frac{\partial}{\partial\varepsilon} \delta V
+ \alpha B^+ B - i B^+ \bar D^2 V - i B D^2 V \Bigg),\qquad
\end{eqnarray}

\noindent
which is invariant under the BRST transformations:

\begin{eqnarray}\label{SUSY_BRST_Transformations}
&& \delta V = \varepsilon\Bigg[- \frac{i}{2}
\mbox{\boldmath$V$}\mbox{ctg}\mbox{\boldmath$V$}\Big(c-c^+\Big)
+\frac{1}{2}\mbox{\boldmath$V$}\Big(c+c^+\Big)\Bigg];
\nonumber\\
&& \delta B = 0;\qquad\qquad\qquad\quad
\delta B^+ = 0;\vphantom{\Big(}\nonumber\\
&& \delta\tilde c = i\varepsilon B;\vphantom{\Big(}
\qquad\qquad\qquad\
\delta\tilde c^+ = - i\varepsilon B^+;\vphantom{\Big(}\nonumber\\
&& \delta c = -\varepsilon c^2;\vphantom{\Big(}
\qquad\qquad\qquad
\delta c^+ = - \varepsilon (c^+)^2.\vphantom{\Big(}
\end{eqnarray}

\noindent
Here we use the following notations:

\begin{equation}
V = -ie V^a t^a,\qquad \mbox{tr}(t^a t^b) = \frac{1}{2} \delta^{ab},
\end{equation}

\noindent
where $t^a$ are hermitian generators, so that $V^a$ are real scalar
superfields and for any function $f(V) = c_0 + c_1 V + c_2 V^2 +\ldots$
we define

\begin{equation}
f(\mbox{\boldmath$V$}) A \equiv c_0 A + c_1 [V,A] + c_2 [V,[V,A]] +\ldots.
\end{equation}

\noindent
$W_a$ is a chiral spinor superfield, given by the equation

\begin{equation}\label{W_Definition}
W_a \equiv \frac{1}{32} \bar D (1-\gamma_5) D \Big(e^{-2iV}
(1+\gamma_5)D_a e^{2iV}\Big),
\end{equation}

\noindent
where $D$ is a supersymmetric covariant derivative:

\begin{equation}
D = \frac{\partial}{\partial\bar\theta} - i\gamma^\mu\theta\,\partial_\mu.
\end{equation}

It is convenient to introduce the generating functional with additional
sources $G$ and $g$:

\begin{eqnarray}\label{Z}
&& Z = \int d\mu\,\exp\Bigg\{i S
+ i\,\mbox{tr}\int d^4x\,d^4\theta\,\Big(V J
+ G\,\frac{\partial}{\partial\varepsilon}\delta V\Big)
+ i\,\mbox{tr}\int d^4x\,d^2\theta
\times\nonumber\\
&& \times
\Big(j_c\,c
+ \bar j_c \bar c
+ g \frac{\partial}{\partial\varepsilon}\delta c \Big)
+ i\,\mbox{tr}\int d^4x\,d^2\bar\theta\,\Big(j_c^+\,c^+
+ \bar j_c^+ \bar c^+
+ g^+ \frac{\partial}{\partial\varepsilon}\delta c^+ \Big)\Bigg\}.\qquad
\end{eqnarray}

\noindent
Starting from this generating functional we construct the generating
functional for connected Green functions

\begin{equation}
W = - i\ln Z
\end{equation}

\noindent
and the effective action

\begin{eqnarray}
\Gamma = W - \int d^4x\,d^4\theta\, V J
- \int d^4x\,d^2\theta\,\Big(j_c c + \bar j_c\bar c\Big)
- \int d^4x\,d^2\bar\theta\,\Big(j_c^+ c^+ + \bar j_c^+\bar c^+\Big),
\end{eqnarray}

\noindent
where the sources should be expressed via equations

\begin{equation}
V = \frac{\delta W}{\delta J};\qquad
c = \frac{\delta W}{\delta j_c};\qquad
\bar c = \frac{\delta W}{\delta \bar j_c};\qquad
c^+ = \frac{\delta W}{\delta j_c^+};\qquad
\bar c^+ = \frac{\delta W}{\delta \bar j_c^+}.
\end{equation}

\noindent
Note, that the effective action will depend on the sources $G$ and $g$
as on parameters.


\section{Invariant renormalization}
\hspace{\parindent}
\label{Section_Universal_Renormalization}

Performing in the generating functional (\ref{Z}) a substitution
(\ref{SUSY_BRST_Transformations}), it is easy to derive STI, which
can be written as

\begin{eqnarray}
&& \mbox{tr}\int d^4x\,d^4\theta\,\Bigg[
\frac{\delta\Gamma}{\delta V} \frac{\delta\Gamma}{\delta G}
+ \frac{2}{\alpha}\Bigg(D^2 V \frac{\delta\Gamma}{\delta\bar c}
- \bar D^2 \frac{\delta\Gamma}{\delta \bar c^+} \Bigg) \Bigg]
+\nonumber\\
&& \qquad\qquad\qquad\qquad\qquad
+ \mbox{tr} \int d^4x\,d^2\theta\,\frac{\delta\Gamma}{\delta c}
\frac{\delta \Gamma}{\delta g} + \mbox{tr}\int d^4x\,d^2\bar\theta\,
\frac{\delta\Gamma}{\delta c^+} \frac{\delta \Gamma}{\delta g^+} = 0.
\qquad
\end{eqnarray}

Due to invariance of the generating functional under the transformations
$\bar c \to \bar c + \epsilon(x)$, where $\epsilon(x)$ is an
arbitrary anticommuting function, the following identities take place:

\begin{equation}
\Big\langle \frac{1}{32 e^2}\bar D^2
\frac{\partial}{\partial\varepsilon}\delta V
+ \bar j_c \Big\rangle = 0;\qquad
\Big\langle \frac{1}{32 e^2} D^2
\frac{\partial}{\partial\varepsilon}\delta V
+ \bar j_c^+ \Big\rangle = 0.
\end{equation}

%
%
%

\noindent
In order to write these STI in terms of the effective action it is
convenient to define a functional $\tilde\Gamma$ by subtracting the
gauge fixing term from $\Gamma$:

\begin{equation}
\tilde\Gamma \equiv \Gamma - \frac{1}{32 e^2 \alpha} \mbox{tr}
\int d^4x\,d^4\theta\, V \Big(D^2\bar D^2 + \bar D^2 D^2\Big) V.
\end{equation}

\noindent
This functional satisfies the following equations:

\begin{eqnarray}\label{Main_ST_Identity}
&& \mbox{tr}\int d^4x\,d^4\theta\,
\frac{\delta\tilde \Gamma}{\delta V}
\frac{\delta\tilde\Gamma}{\delta G}
+ \mbox{tr} \int d^4x\,d^2\theta\,
\frac{\delta\tilde \Gamma}{\delta c}
\frac{\delta \tilde\Gamma}{\delta g}
+ \mbox{tr}\int d^4x\,d^2\bar\theta\,
\frac{\delta\tilde\Gamma}{\delta c^+}
\frac{\delta \tilde\Gamma}{\delta g^+} = 0;\\
\label{Ghost_ST_Identities}
&& - \frac{1}{32 e^2} \bar D^2 \frac{\delta\tilde\Gamma}{\delta G}
+ \frac{\delta\tilde\Gamma}{\delta \bar c} = 0;\qquad
- \frac{1}{32 e^2} D^2 \frac{\delta\tilde\Gamma}{\delta G}
+ \frac{\delta\tilde\Gamma}{\delta \bar c^+} = 0.
\end{eqnarray}

\noindent
Note, that up to now we did not set any field equal to 0. Moreover,
identity (\ref{Main_ST_Identity}) is not linear. So, to treat these
identities the following algorithm should be used:

1. In order to find relations between different Green functions it
is necessary to differentiate identities (\ref{Main_ST_Identity}) and
(\ref{Ghost_ST_Identities}) with respect to the arguments and then set
all fields equal to 0.

2. Identity (\ref{Main_ST_Identity}) should be expanded over the Plank
constant $\hbar$, that corresponds to the loop expansion. Let us present
the effective action as follows:

\begin{equation}
\tilde\Gamma = S + \hbar \Delta \Gamma^{(1)}
+ \hbar^2 \Delta\Gamma^{(2)} + \ldots
\end{equation}

\noindent
Then in the lowest order this identity is satisfied due to BRST-invariance
of the classical action. At the arbitrary order one has
\begin{equation}
\Delta \Gamma^{(n)} \cdot S+S \cdot \Delta \Gamma^{(n)}
= - \sum_{m=1}^{n-1} \Delta \Gamma^{(m-n)} \cdot \Delta \Gamma^{(m)}
\label{1*}
\end{equation}

\noindent
where $\cdot$ denotes an operator, which appears after differentiation
of (\ref{Main_ST_Identity}):

\begin{eqnarray}\label{Cdot_Operator}
&& A\cdot B \equiv \mbox{tr}\int d^4x\,d^4\theta\,
\Bigg(\frac{\delta A}{\delta V}
\frac{\delta B}{\delta G}
+\frac{\delta B}{\delta V}
\frac{\delta A}{\delta G}\Bigg)
+\nonumber\\
&& + \mbox{tr} \int d^4x\,d^2\theta\,
\Bigg(\frac{\delta A}{\delta c}
\frac{\delta B}{\delta g}
+\frac{\delta B}{\delta c}
\frac{\delta A}{\delta g}\Bigg)
+ \mbox{tr}\int d^4x\,d^2\bar\theta\,
\Bigg(\frac{\delta A}{\delta c^+}
\frac{\delta B}{\delta g^+}
+\frac{\delta B}{\delta c^+}
\frac{\delta A}{\delta g^+}\Bigg) = 0.\qquad
\end{eqnarray}

\noindent
The identity (\ref{1*}) holds if a regularization
preserves gauge invariance. If a noninvariant regularization is used the
invariance may be restored by adding
local counterterms to the proper vertices. This is the procedure adopted
in the algebraic renormalization scheme, where the main problem is to
find the renormalized action $S_{ren}$, providing the effective action
which satisfies ST identities. As the explicit determination of all
noninvariant counterterms is a difficult problem, we adopt another
approach, which allows to write directly the gauge invariant renormalized
effective action by solving ST identities.

Our goal is to get the renormalized effective action $\tilde\Gamma_R$,
which satisfies STI by construction:

\begin{equation}
\tilde\Gamma_R = S_{ren} + \hbar \Delta\gamma^{(1)}
+ \hbar^2\Delta\gamma^{(2)}+\ldots.
\end{equation}

\noindent
Here $\Delta\gamma^{(i)}$ are some functions, which are constructed from
$\Delta\Gamma^{(i)}$ so that the renormalized effective action satisfies
renormalized STI. The renormalized action is given by

\begin{equation}
S_{ren} = S - \hbar \Delta\gamma^{(1)}_{div}
- \hbar^2\Delta\gamma^{(2)}_{div} + \ldots,
\end{equation}

\noindent
so that

\begin{equation}\label{Gamma_R1}
\tilde\Gamma_R = S + \hbar \Delta\Gamma_R^{(1)}
+ \hbar^2\Delta\Gamma_R^{(2)}+\ldots,
\end{equation}

\noindent
where

\begin{equation}\label{Gamma_R2}
\Delta\Gamma_R^{(i)} \equiv \Delta\gamma^{(i)}
- \Delta\gamma^{(i)}_{div}.
\end{equation}

If the theory is anomaly free, it is always possible to find such
counterterms $ \Delta \gamma^{(i)}_{div}$, that

\begin{equation}\label{Equality}
\Delta\gamma^{(i)}-\Delta\gamma^{(i)}_{div}
= \Delta\Gamma^{(i)}-\Delta\Gamma^{(i)}_{div},
\end{equation}

\noindent
in the limit, when a regularization is removed.
Here $\Delta\Gamma^{(i)}$ and $\Delta\Gamma_{div}^{(i)}$ correspond
to $\Delta\gamma^{(i)}$ and $\Delta\gamma_{div}^{(i)}$ in case of
invariant regularization. Below we shall show that by solving STI one
may get an  expression for $\Delta \Gamma_R^{(i)}$ without finding
explicitely noninvariant counterterms $\Delta \gamma_{div}^{(i)}$.

We shall use the eq.(\ref{1*}) to determine the renormalized Green function of
order $n$ assuming that according to $R$-operation the subtractions in all
divergent subgraphs, corresponding to $\Gamma^{(m)},\ m<n$ are done in
agreement with the eq.(\ref{1*}).

We shall work with the equation (\ref{1*}) assuming that it is
differentiated with respect to the fields, and then all the fields are
put equal to zero. In this way we get a relation which expresses an
arbitrary Green function with at least one chiral external gauge line
in terms of other Green functions which are of the same order in $h$
but have one external gauge line less and lower order Green functions.
Symbolically

\begin{equation}
D_x^2 \Gamma^n(x...)= R(\Gamma^m)
\label{2*}
\end{equation}

\noindent
where $R(\Gamma_m)$ denotes a product of lower order Green functions
and the Green functions of the same order but having one external gauge
line less. We assume that all Green functions of order $m<n$ are
renormalized in agreement with the eq.(\ref{1*}) and accordingly the
divergent subgraphs in the function $\Gamma_n$ are subtracted following
the R-operation. Then the eq.(\ref{2*}) fixes the overall subtraction
in the function $\Gamma^n$ so that the renormalized function satisfies
STI. Indeed let us present the function $\Gamma_n$ in the form

\begin{equation}\label{Gamma_Equation}
\Gamma_n(\theta,p)= \sum_i B_i(\theta,p)F_i(p)
\label{3*}
\end{equation}

\noindent
where the functions $B_i$ are polynomials in $\theta,p$ and $F_i$ depend
only on $p$. The renormalization is achieved by subtracting a polynomial
of $F_i(p)$. We will choose the polynomials $B_i$ so that their chiral
parts

\begin{equation}
Q_i=D_x^2 B_i
\label{4*}
\end{equation}

\noindent
are either linear independent, or equal to zero.

The r.h.s. of eq.(\ref{2*}) may be decomposed over the polynomials $Q_i$

\begin{equation}
R(\Gamma_m)= \sum_i Q_i(\theta,p) R_i(p)
\label{5*}
\end{equation}

\noindent
In practice to find the functions $R_i$ explicitely one has to perform
a multiplication of the coefficients functions of the Green functions
$\Gamma_m$, entering the r.h.s. of eq.(\ref{2*}) using the supersymmetry
algebra.

According to our assumption all functions at the r.h.s. of eq.(\ref{2*})
are finite. Using linear independence of coefficient functions $Q_i$, we
find that  the renormalized function $\Gamma_n$  may be written as follows

\begin{equation}
\Gamma_n^R= \sum_i B_i(\theta,p)R_i(p)+ \sum_j B_j(\theta,p)F_j(p)
\label{6*}
\end{equation}

\noindent
where the second sum runs over $j$ for which $Q_j=0$.
It completes the renormalization.

Such renormalization can be made in the following sequence:

At the first step we should calculate one-loop correction to the
two-point Green functions of the ghost fields:

\begin{equation}
\frac{\delta^2}{\delta c^+ \delta \bar c}\Delta\Gamma^{(1)}_R;\qquad
\frac{\delta^2}{\delta c\,\delta \bar c^+}\Delta\Gamma^{(1)}_R.
\end{equation}

\noindent
STI do not impose any restrictions on the renormalization of these
functions. Therefore, they can be renormalized in arbitrary way, say, by
subtraction at some fixed normalization point $\mu_c$. At the next step
we calculate functions

\begin{equation}\label{GC_Functions}
\frac{\delta^2}{\delta c^+ \delta G}\Delta\Gamma^{(1)}_R;\qquad
\frac{\delta^2}{\delta c\,\delta G}\Delta\Gamma^{(1)}_R.
\end{equation}

\noindent
For this purpose it is necessary to use STI, which is obtained by
differentiating equation (\ref{Ghost_ST_Identities}) over $c$.
Then we proceed to the renormalization of the function

\begin{equation}\label{GCC_Function}
\frac{\delta^3 \Delta\Gamma^{(1)}}{\delta c\,\delta\bar c\,\delta V},
\end{equation}

\noindent
using the corresponding STI. Having constructed functions
(\ref{GC_Functions}) and (\ref{GCC_Function}) it is possible to
renormalize the function

\begin{equation}
\frac{\delta^3 \Delta\Gamma^{(1)}}{\delta c^+\,\delta\bar c\,\delta V}.
\end{equation}

\noindent
In each case the STI, which are needed for the renormalization, are
obtained by differentiating of (\ref{Main_ST_Identity}) or
(\ref{Ghost_ST_Identities}). Then we differentiate all STI over $V$ and
proceed to the renormalization of functions with larger powers of $V$,
starting from

\begin{equation}
\frac{\delta^3 \Delta\Gamma^{(1)}}{\delta c\,\delta G\,\delta V}.
\end{equation}

Renormalization of the two-point Green function for gauge field is made
similar to the QED case \cite{SlSt}: If we introduce the function $\Pi$

\begin{eqnarray}\label{Definition_Of_Pi}
&& (2\pi)^4 \delta^4\Big((p_1)_\mu + (p_2)_\mu\Big)\,
\Pi \Big[(\theta_{x_1},p_1),(\theta_{x_2},-p_1)\Big]
\equiv \qquad\vphantom{\Bigg(}\nonumber\\
&& \equiv
\int d^4x_1 d^4x_2\,
\frac{\delta^{2}\tilde\Gamma}{\delta V_{x_1} \delta V_{x_2}}
\Bigg|_{\phi,\tilde\phi,V=0} \exp\Big(i (p_1)_\mu x_1^\mu
+ i (p_2)_\mu x_2^\mu \Big),
\end{eqnarray}

\noindent
then

\begin{eqnarray}\label{Original_Pi2}
&& \Pi\Big[(\theta_x,p),(\theta_y,-p)\Big]
= F_1(p)\,p^2\,\Pi_{1/2}\,\delta^4(\theta_x-\theta_y)
+ F_2(p)\,\delta^4(\theta_x-\theta_y),\qquad
\end{eqnarray}

\noindent
where

\begin{equation}
\Pi_{1/2} = - \frac{1}{16 \partial^2} D^a \bar D^2 C_{ab} D^b.
\end{equation}

\noindent
In this case

\begin{eqnarray}
B_1(\theta,p) = p^2 \Pi_{1/2}\,\delta^4(\theta_x-\theta_y);\qquad
B_2(\theta,p) = \delta^4(\theta_x-\theta_y)
\end{eqnarray}

The renormalized Green functions are obtained by subtracting
from $F_i(p)$ some polynomials chosen to provide STI for the function
$\Pi^r$

\begin{equation}\label{Renormalized_Pi}
\Pi^r[\theta,p] = \sum\limits_i B_i(\theta,p) \Big(F_i(p) - P_i(p)\Big),
\end{equation}

\noindent
For the two-point Green function substitution of
(\ref{Original_Pi2}) into the STI

\begin{equation}
\bar D^2_y\, \frac{\delta^2\Delta\Gamma^{(1)}}{\delta V_z \delta V_y} = 0,
\end{equation}

\noindent
which is obtained by differentiating the linearized STI
(\ref{Main_ST_Identity}) over $c^+$ and $V$, gives the following equation:

\begin{equation}
\Big(F_2(p) - P_2(p)\Big)\,
D^2_x\delta^4(\theta_x-\theta_y)=0.
\end{equation}

\noindent
Therefore,

\begin{equation}
P_2(p) = F_2(p),
\end{equation}

\noindent
while the function $P_1$ can not be defined from STI and corresponds to
a gauge invariant counterterm. It is convenient to choose

\begin{equation}
P_1(p) = F_1(\mu_\pi),
\end{equation}

\noindent
where $\mu_\pi$ is a normalization point. Then the renormalized two-point
Green function can be written as

\begin{equation}\label{Pi2_Subtraction}
\Pi^r\Big[(\theta_x,p),(\theta_y,-p)\Big] =
\Big(F_1(p) - F_1(\mu_\pi)\Big)\,p^2\,\Pi_{1/2}\,
\delta^4(\theta_x-\theta_y).\qquad
\end{equation}

\noindent
This Green function satisfies the equation

\begin{equation}
D^2_x \Pi^r\Big[(\theta_x,p),(\theta_y,-p)\Big] =0,
\end{equation}

\noindent
which is a supersymmetric generalization of transversality condition
in the usual quantum electrodynamics.

Renormalization of the three-point function for the gauge field can
be made according to the above algorithm using the corresponding STI.

The procedure described above may be interpreted as finding local
counterterms, which can be written as

\begin{eqnarray}\label{Delta_S}
&& \Delta S =
\frac{1}{16 e^2}\mbox{tr}\int d^4x\,d^4\theta\,\Big(\bar c^+-\bar c\Big)
\Bigg(i\sum\limits_{n=1}^\infty a_n \mbox{\boldmath$V$}^n (c-c^+)
+ \sum\limits_{n=0, \ne 1}^\infty b_n
\mbox{\boldmath$V$}^n (c+c^+) \Bigg)
+\nonumber\\
&& + \mbox{tr}\int d^4x\,d^4\theta\,G \Bigg(
i\sum\limits_{n=0}^\infty f_n \mbox{\boldmath$V$}^n (c-c^+)
+ \sum\limits_{n=0}^\infty h_n \mbox{\boldmath$V$}^n (c+c^+) \Bigg)
+ \mbox{tr}\int d^4x\,d^4\theta\,F[V].\qquad
\end{eqnarray}

\noindent
After adding these counterterms the renormalized effective action
$\Delta\Gamma^{(1)}_R$ will satisfy renormalized STI. However, the
above algorithm allows to avoid writing and tuning of all terms in
(\ref{Delta_S}), that may cause considerable technical difficulties.


\section{One-loop renormalization with the simplest PV-regularization}
\hspace{\parindent}
\label{Section_Example}

In order to illustrate the scheme, presented above, we consider
$N=1$ supersymmetric Yang-Mills theory and calculate its $\beta$-function
using regularization by (usual) higher derivatives, complemented by
Pauli-Villars fields for regularization of one-loop divergences.

The regularization is introduced by adding the higher derivative term,
so that the regularized action is equal to

\begin{eqnarray}
&& S_{reg} = \frac{1}{2 e^2}\mbox{Re}\,\mbox{tr}\int d^4x\ d^2\theta\
W_a C^{ab} \Bigg(1 + \frac{\partial^{2n}}{\Lambda^{2n}} \Bigg) W_b
+\nonumber\\
&& \qquad\quad
+ \frac{1}{16 e^2} \mbox{tr} \int d^4x\,d^4\theta \Bigg(
\Big(\bar c^+ -\bar c\Big)\,\frac{\partial}{\partial\varepsilon} \delta V
+ \alpha B^+ B - i B^+ \bar D^2 V - i B D^2 V \Bigg),\qquad
\end{eqnarray}

\noindent
However, the higher derivative term does not regularize one-loop
divergences. Therefore, it is necessary to add in the generating
functional PV-fields, that can be made by the following way:

\begin{eqnarray}\label{Modified_Z}
&& Z = \int d\mu\,
\prod\limits_i \Big(\det PV(V,M_i)\Big)^{\alpha_i}
\prod\limits_i \Big(\det pv(V,m_i)\Big)^{\beta_i}
\times\nonumber\\
&& \times
\exp\Bigg\{i S_{reg}
+ i\,\mbox{tr}\int d^4x\,d^4\theta\,\Big(V J
+ G\,\frac{\partial}{\partial\varepsilon}\delta V\Big)
+ i\,\mbox{tr}\int d^4x\,d^2\theta
\times\nonumber\\
&& \times
\Big(j_c\,c
+ \bar j_c \bar c
+ g \frac{\partial}{\partial\varepsilon}\delta c \Big)
+ i\,\mbox{tr}\int d^4x\,d^2\bar\theta\,\Big(j_c^+\,c^+
+ \bar j_c^+ \bar c^+
+ g^+ \frac{\partial}{\partial\varepsilon}\delta c^+ \Big)\Bigg\}.\qquad
\end{eqnarray}

\noindent
where

\begin{eqnarray}\label{PV_Determinants}
&& \Big(\det PV(V,M)\Big)^{-1}
=\nonumber\\
&& = \int DW
\exp\Bigg\{\frac{i}{2} \int d^4x\,d^4\theta_x\,d^4y\,d^4\theta_y\,
W_x^a \frac{\delta^2 S_{reg}}{\delta V^a_x\delta V^b_y} W^b_y
- \frac{i}{4} M^2 \int d^4x\,d^4\theta\,W_a^2 \Bigg\},
\qquad\nonumber\\
&& \Big(\det pv(V,m)\Big)^{-1}
=\nonumber\\
&& = \int DC\,D\bar C
\exp\Bigg\{
\frac{i}{16 e^2} \mbox{tr} \int d^4x\,d^4\theta \Bigg(
\Big(\bar C^+ -\bar C\Big)\,\frac{\partial}{\partial\varepsilon}
\delta V(V,C,C^+)
+\nonumber\\
&& \qquad\qquad\qquad\qquad\qquad\quad\ \
+ \frac{i m}{2 e^2} \mbox{tr}
\int d^4x\,d^2\theta\, \bar C\,C
+ \frac{im}{2e^2}
\mbox{tr}\int d^4x\,d^2\bar\theta\,\bar C^+ C^+ \Bigg\},\quad
\end{eqnarray}

\noindent
and the coefficients $\alpha_i$ and $\beta_i$ satisfy the following
equations:

\begin{eqnarray}
&& \sum\limits_i \alpha_i = 1;\qquad \sum\limits_i \alpha_i M_i^2 = 0;
\nonumber\\
&& \sum\limits_i \beta_i = 1;\qquad \sum\limits_i \beta_i m_i^2 = 0.
\end{eqnarray}

\noindent
Here $W^a$, $C$ and $\bar C$ are PV fields, $M_i$ are masses of PV-fields,
corresponding to $V$ superfield and $m_i$ are masses of PV-fields,
corresponding to ghosts. We will assume, that all $M_i$ and $m_i$ are
proportional to the parameter $\Lambda$ in the higher derivative term.

\begin{figure}[h]
\hspace*{7mm}
\epsfxsize13.5truecm\epsfbox{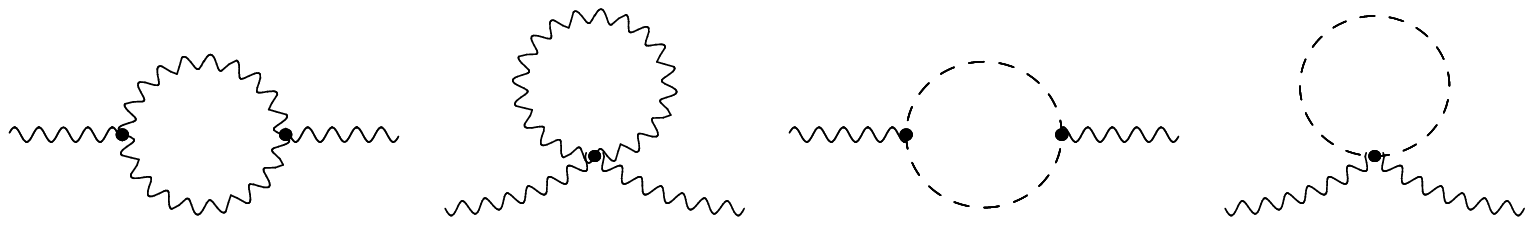}
\caption{One-loop contribution to the two-point Green function of the
gauge superfield.}
\label{Two_Point_Diagrams}
\end{figure}

In order to find two-point Green function for the gauge superfield
it is necessary to calculate Feynman diagrams, presented at Figure
\ref{Two_Point_Diagrams}. The result appeared to be

\begin{eqnarray}\label{Two_Point_Function}
&& \Delta\Gamma^{(1)}_{2-point}
= -i e^2 c_1 \int\frac{d^4p}{(2\pi)^4}\, V^a(-p)\,\partial^2 \Pi_{1/2} V^a(p)
\times\nonumber\\
&& \times
\Bigg\{\frac{1}{2}\Big(1+(-1)^n p^{2n}/\Lambda^{2n}\Big)
\int\frac{d^4k}{(2\pi)^4} \Bigg(\frac{1}{k^2 (k+p)^2
\Big(1+(-1)^n k^{2n}/\Lambda^{2n}\Big)}
-\nonumber\\
&& - \sum\limits_k \alpha_k
\frac{1}{(k^2-M_j^2) \Big((k+p)^2-M_j^2\Big)
\Big(1+(-1)^n k^{2n}/\Lambda^{2n}\Big)}\Bigg)
+\nonumber\\
&& +\frac{1}{4}
\int\frac{d^4k}{(2\pi)^4} \Bigg(\frac{1}{k^2 (k+p)^2}
- \sum\limits_k \alpha_k
\int\frac{d^4k}{(2\pi)^4}
\frac{1}{(k^2-M_j^2) \Big((k+p)^2-M_j^2\Big)}\Bigg)\Bigg\}
+\nonumber\\
&& + e^2 c_1 \int\frac{d^4p}{(2\pi)^4}\,V_a(-p)\, V_a(p)
\times\vphantom{\Bigg(}\nonumber\\
&&\times
\Bigg\{
\sum\limits_j \frac{i\alpha_j}{3} \int\frac{d^4k}{(2\pi)^4}
\frac{1}{(k^2-M_j^2)}
-\sum\limits_j\frac{i\alpha_j}{8} \int\frac{d^4k}{(2\pi)^4}
\frac{2(k+p)^2+p^2}{
(k^2-M_j^2)\Big((k+p)^2-M_j^2\Big)}
+\nonumber\\
&& +\sum\limits_j\frac{i\alpha_j}{4} \int\frac{d^4k}{(2\pi)^4}
\frac{M_j^2\Big(1+(-1)^n(k+p)^{2n}/\Lambda^{2n}\Big)}{
(k^2-M_j^2)\Big(1+(-1)^n k^{2n}/\Lambda^{2n}\Big)\Big((k+p)^2-M_j^2\Big)}
-\nonumber\\
&& -\sum\limits_j \frac{i\beta_j}{12} \int\frac{d^4k}{(2\pi)^4}
\frac{1}{(k^2-m_j^2)}
+\sum\limits_j \frac{i\beta_j}{8} \int\frac{d^4k}{(2\pi)^4}
\frac{p^2}{(k^2-m_j^2) \Big((k+p)^2-m_j^2\Big)}
\Bigg\}\qquad
\end{eqnarray}

\noindent
Substituting this expression to equation (\ref{Definition_Of_Pi}) we
obtain the function $\Pi$, which is equal to (\ref{Original_Pi2}) with

\begin{eqnarray}
&& F_1(p)
= i e^2 c_1
\Bigg\{\Big(1+(-1)^n p^{2n}/\Lambda^{2n}\Big)
\int\frac{d^4k}{(2\pi)^4} \Bigg(\frac{1}{k^2 (k+p)^2
\Big(1+(-1)^n k^{2n}/\Lambda^{2n}\Big)}
-\nonumber\\
&& - \sum\limits_k \alpha_k
\frac{1}{(k^2-M_j^2) \Big((k+p)^2-M_j^2\Big)
\Big(1+(-1)^n k^{2n}/\Lambda^{2n}\Big)}\Bigg)
+\nonumber\\
&& +\frac{1}{2}
\int\frac{d^4k}{(2\pi)^4} \Bigg(\frac{1}{k^2 (k+p)^2}
- \sum\limits_k \alpha_k
\int\frac{d^4k}{(2\pi)^4}
\frac{1}{(k^2-M_j^2) \Big((k+p)^2-M_j^2\Big)}\Bigg)\Bigg\};
\nonumber\\
&& F_2(p) = 2 e^2 c_1 \Bigg\{
\sum\limits_j \frac{i\alpha_j}{3} \int\frac{d^4k}{(2\pi)^4}
\frac{1}{(k^2-M_j^2)}
-\nonumber\\
&&\qquad\qquad\qquad\qquad\qquad
-\sum\limits_j\frac{i\alpha_j}{8} \int\frac{d^4k}{(2\pi)^4}
\frac{2(k+p)^2+p^2}{
(k^2-M_j^2)\Big((k+p)^2-M_j^2\Big)}
+\nonumber\\
&& +\sum\limits_j\frac{i\alpha_j}{4} \int\frac{d^4k}{(2\pi)^4}
\frac{M_j^2\Big(1+(-1)^n(k+p)^{2n}/\Lambda^{2n}\Big)}{
(k^2-M_j^2)\Big(1+(-1)^n k^{2n}/\Lambda^{2n}\Big)\Big((k+p)^2-M_j^2\Big)}
-\nonumber\\
&& -\sum\limits_j \frac{i\beta_j}{12} \int\frac{d^4k}{(2\pi)^4}
\frac{1}{(k^2-m_j^2)}
+\sum\limits_j \frac{i\beta_j}{8} \int\frac{d^4k}{(2\pi)^4}
\frac{p^2}{(k^2-m_j^2) \Big((k+p)^2-m_j^2\Big)}
\Bigg\}\qquad
\end{eqnarray}

In order to calculate the integrals in this expression, first it is
necessary to perform the Wick rotation. Then it is easy to see, that

\begin{eqnarray}\label{Integrals_DK_Integral}
F_1(p) = - \frac{3}{16} e^2 c_1
\Big(\ln \frac{\Lambda}{p} + \mbox{finite terms}\Big)
\end{eqnarray}

\noindent
It is easy to verify, that the function $F_2$ is finite and proportional
to $p^2$. After taking a limit $\Lambda\to\infty$ (and therefore,
$M_j\to\infty$ and $m_j\to\infty$) we obtain, that

\begin{eqnarray}\label{Final_Two_Point_Function}
&& \Pi\Big[(\theta_x,p),(\theta_y,-p)\Big]
= -\frac{3}{16\pi^2} e^2 c_1
p^2 \Pi_{1/2} \delta^4(\theta_x-\theta_y)
\Big(\ln \frac{\Lambda}{p} + \mbox{finite terms}\Big)
+\nonumber\\
&& + e^2 c_1 \delta^4(\theta_x-\theta_y)
p^2 C\Big(\alpha_j,\beta_j,M_j/\Lambda,m_j/\Lambda\Big)
\Bigg\}.\qquad
\end{eqnarray}

\noindent
where the constant $C$ is given by

\begin{equation}
p^2 C=\lim\limits_{\Lambda\to\infty} F_2(p)
\end{equation}

\noindent
and in the general case is not 0, that can be easily verified. Then
the renormalized function $\Pi^r$, obtained according to prescription
(\ref{Pi2_Subtraction}), is

\begin{equation}
\Pi\Big[(\theta_x,p),(\theta_y,-p)\Big]
= -\frac{3}{16\pi^2} e^2 c_1
p^2 \Pi_{1/2} \delta^4(\theta_x-\theta_y)
\Big(\ln \frac{\mu}{p} + \mbox{finite terms}\Big)
\end{equation}

\noindent
and corresponds to the following renormalized contribution to the
effective action:

\begin{equation}
\Delta\Gamma^{(1)}_R
= \frac{3 e^2}{32\pi^2} c_1 \int\frac{d^4p}{(2\pi)^4}\,
V^a(-p)\,\partial^2 \Pi_{1/2} V^a(p)\,\Big(\ln\frac{\mu}{p}
+\mbox{finite terms}\Big).
\end{equation}

\begin{figure}[h]
\epsfxsize15.0truecm\epsfbox{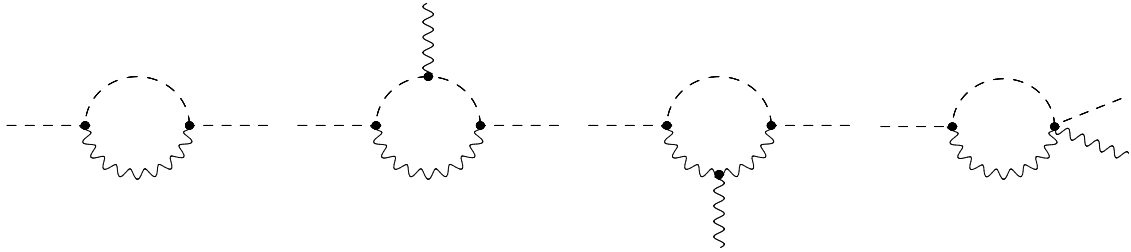}
\caption{One-loop contributions to the ghost propagator and ghost-gluon
vertex.}
\label{Ghost_Diagrams}
\end{figure}

One-loop correction to the ghost propagator corresponding to the first
diagram, presented in Fig. \ref{Ghost_Diagrams} is found to be zero. We
also verified, that the one-loop ghost-gluon vertex, corresponding to
the other diagrams in this figure, is finite. Therefore, the one-loop
$\beta$-function

\begin{equation}\label{Beta_Definition}
\beta(\alpha) \equiv \frac{d}{d\ln\mu}\Bigg(\frac{e^2}{4\pi}\Bigg),
\end{equation}

\noindent
calculated in the noninvariant regularization, described above is

\begin{equation}
\beta(\alpha) = -\frac{3\alpha^2 c_1}{2\pi} + O(\alpha^3)
\end{equation}

\noindent
and agrees with the calculations made by DRED. However, it is
interesting to find the two-loop contribution to the $\beta$-function
(\ref{Beta_Definition}) in the considered regularization, because
in SUSY QED \cite{hep,HD_And_DRED} such contribution is different
from the DRED result and is equal to zero.


\section{Conclusion.}
\hspace{\parindent}

In this paper we presented a renormalization procedure for SUSY
Yang-Mills theories, which guarantees BRST-invariance of the renormalized
theory for any intermediate regularization. SUSY Slavnov-Taylor identities
are incorporated into subtractions, which allows to avoid the appearance
of noninvariant counterterms. The proposed procedure is illustrated by
the calculation of the one-loop $\beta$-function for the $N=1$ SUSY
Yang-Mills theory with the noninvariant version of higher derivative
regularization. However it would be interesting to calculate two-loop
quantum corrections and compare them with the correspondent result in
SUSY QED \cite{hep,HD_And_DRED}.


\bigskip
\bigskip

\noindent
{\Large\bf Acknowledgments}

\bigskip

This work was partially supported by RBRF grant N 02-01-00126 and by the
grant for support of leading scientific schools.


\pagebreak

\end{document}